\documentclass[sigconf,nonacm]{acmart}
\usepackage{soul} %
\usepackage{tcolorbox} %
\usepackage{pifont} %
\usepackage{makecell} %

\newtcolorbox{answerbox}{
boxrule=0.5pt,
left=3pt, right=3pt, top=3pt, bottom=3pt
}

\newcommand*\blackcircled[1]{%
    \tikz[baseline=(char.base)]{
        \node[
            shape=circle, 
            draw=black,          %
            fill=black,          %
            text=white,          %
            inner sep=0pt,       %
            font=\bfseries\small,%
            minimum size=8       %
        ] (char) {#1};
    }%
}

\newcommand{\ourproj}[1]{\emph{XuanJia}}

\begin{document}

\title{
XuanJia: A Comprehensive Virtualization-Based Code Obfuscator for Binary Protection
}

\author{Xianyu Zou}
\email{2012077@mail.nankai.edu.cn}
\affiliation{%
  \department{College of Computer Science}
  \institution{Nankai University}
  \city{Tianjin}
  \country{China}
}

\author{Xiaoli Gong}
\email{gongxiaoli@nankai.edu.cn}
\affiliation{%
  \department{College of Computer Science}
  \institution{Nankai University}
  \city{Tianjin}
  \country{China}
}

\author{Jin Zhang}
\email{nkzhangjin@nankai.edu.cn}
\affiliation{%
  \department{College of Computer Science}
  \institution{Nankai University}
  \city{Tianjin}
  \country{China}
}

\author{Shiyang Li}
\email{li004074@umn.edu}
\affiliation{%
  \department{Department of Computer Science and Engineering}
  \institution{University of Minnesota at Twin Cities}
  \city{Minneapolis}
  \state{MN}
  \country{USA}
}

\author{Pen-Chung Yew}
\email{yew@umn.edu}
\affiliation{%
  \department{Department of Computer Science and Engineering}
  \institution{University of Minnesota at Twin Cities}
  \city{Minneapolis}
  \state{MN}
  \country{USA}
}

\begin{abstract}

Virtualization-based binary obfuscation is widely adopted to protect software intellectual property, yet existing approaches leave exception-handling (EH) metadata unprotected to preserve ABI compatibility. 
This exposed metadata leaks rich structural information—such as stack layouts, control-flow boundaries, and object lifetimes—which can be exploited to facilitate reverse engineering. 
In this paper, we present \ourproj{}, a comprehensive VM-based binary obfuscation framework that provides end-to-end protection for both executable code and exception-handling semantics.
At the core of \ourproj{} is ABI-Compliant EH Shadowing, a novel exception-aware protection mechanism that preserves compatibility with unmodified operating system runtimes while eliminating static EH metadata leakage. 
\ourproj{} replaces native EH metadata with ABI-compliant shadow unwind information to satisfy OS-driven unwinding, and securely redirects exception handling to a protected virtual machine where the genuine EH semantics are decrypted, reversed, and replayed using obfuscated code.

We implement \ourproj{} from scratch, supporting 385 x86 instruction encodings and 155 VM handler templates, and design it as an extensible research testbed. 
We evaluate \ourproj{} across correctness, resilience, and performance dimensions. 
Our results show that \ourproj{} preserves semantic equivalence under extensive dynamic and symbolic testing, effectively disrupts automated reverse-engineering tools such as IDA Pro, and incurs negligible space overhead and modest runtime overhead. 
These results demonstrate that \ourproj{} achieves strong protection of exception-handling logic without sacrificing correctness or practicality.

\end{abstract}

\maketitle

\section{Introduction}

Protecting software intellectual property is a long-standing and fundamental challenge in computer security~\cite{collberg2011toward}. 
Although source code protection is typically regarded as the first line of defense, the rapid advancement of reverse-engineering tools has significantly increased the exposure of deployed binaries~\cite{sharif2009automatic, schwartz2010all, yadegari2015symbolic, blazytko2017syntia}. 
To mitigate these threats, various software protection techniques have been proposed following the taxonomy by Collberg~\cite{collberg1997taxonomy}, such as data obfuscation~\cite{linn2003obfuscation}, control-flow flattening~\cite{laszlo2009obfuscating}, and opaque predicates~\cite{collberg1998manufacturing}.

Virtualization obfuscation is a widely adopted technique for protecting software at the binary level~\cite{banescu2016code, rolles2009unpacking}.  
In general, virtualization obfuscation transforms the native binary code of a target function into a customized and opaque intermediate representation, commonly referred to as VM bytecode~\cite{li2022chosen}.
At the same time, a dedicated virtual machine (VM) interpreter is embedded into the protected program; this interpreter fetches, decodes, and executes the bytecode to faithfully emulate the semantics of the original code.
Because the VM interpreter is implemented entirely in software, sophisticated and frequently changing encoding or encryption schemes can be applied to protect the VM bytecode, thereby substantially increasing the difficulty of static and dynamic reverse engineering~\cite{zhang2025inspecting}. 
As a result, virtualization obfuscation has become a core defense mechanism in many industrial-grade commercial obfuscators, including VMProtect~\cite{vmprotect}, Code Virtualizer~\cite{codevirtualizer}, and Obsidium~\cite{obsidium}.

It is essential to preserve semantic equivalence between the protected code and the original implementation, because virtualization-based protection is typically applied to security-critical code regions. 
Replacing the original function body with a VM entry stub is relatively straightforward: during execution, invoking the function transfers control through this stub to the VM interpreter.
However, the function entry point is not the only interface through which the protected code interacts with the surrounding program. 
In particular, language-level exception handling (e.g., C++ exceptions) must be correctly preserved.
Modern software systems and programming languages make extensive use of exception mechanisms~\cite{bradley2019study, priyadarshan2020impact, priyadarshan2023safer}, and failure to faithfully support exception propagation, unwinding, and handling can lead to incorrect program behavior or observable inconsistencies.

Manipulating the exception-handling (EH) process is particularly challenging because it is rigidly governed by the operating system’s Application Binary Interface (ABI). When an exception is raised, the runtime system must unwind the call stack according to the precise location at which the exception is thrown, during which language-specific destructors and cleanup routines are invoked. To support this process, compilers generate function-specific metadata that enables the operating system and language runtime to correctly perform exception propagation and stack unwinding (details are discussed in Section~\ref{sec:exceptionhandling}).
Because this mechanism is directly invoked by the OS runtime and is tightly coupled with the ABI, binary code protection techniques typically leave EH-related components unmodified in order to preserve compatibility and functional correctness. 
For example, Obsidium avoids virtualizing functions that may throw exceptions~\cite{ObsidiumManual}, whereas VMProtect and Code Virtualizer prioritize correctness by retaining the original EH metadata in plaintext~\cite{vmprotect}.

Although this strategy preserves the functional correctness of exception handling, it also exposes critical static information that can be leveraged for binary reverse engineering. 
Based on our analysis (Sections~\ref{sec:exceptionhandling} and~\ref{sec:ehinobfuscatedcode}), we argue that this exposure introduces an exploitable gap in existing protection schemes. 
Specifically, attackers can use EH metadata to identify function entry points and reconstruct the control flow. 
For instance, recent studies~\cite{williams2020egalito, priyadarshan2020impact, pang2021towards} and tools like Ghidra~\cite{Ghidra} and angr~\cite{shoshitaishvili2016sok} use EH metadata to locate function boundaries, treating them as reliable static anchors for code discovery.
Furthermore, advanced decompilers like IDA Pro 9.0 now parse language-specific data (LSData) within the EH metadata to reconstruct C++ \texttt{try-catch} scopes, revealing the function skeleton and control flow that virtualization aims to obscure~\cite{IDAPro9EHparsing}.
Similarly, Ghidra uses LSData to annotate the decompiled code with specific \texttt{try} block ranges and \texttt{catch} handler addresses, explicitly exposing the control flow logic~\cite{GhidraCode}.

In this paper, we propose \ourproj{}, a comprehensive virtualization-based binary obfuscator that secures not only executable code but also exception-handling logic and its associated metadata. Specifically, \ourproj{} introduces an \textbf{ABI-Compliant EH Shadowing} protection mechanism.
During static transformation, \ourproj{} replaces the original exception-handling (EH) metadata with a shadow exception-handling section, thereby eliminating static information leakage while it keeps ABI-compliant and preserves correct interaction with the operating system runtime. 
The original EH metadata is encrypted and relocated into the virtual machine.
At runtime, the shadow handler first participates in the standard unwinding process to maintain ABI compliance, and then transfers control to the protected exception handler within the VM. 
Inside the VM, \ourproj{} reverses the effects of the shadow unwinding and subsequently performs exception propagation and stack unwinding using the decrypted metadata. 
This reimplemented handler is itself protected using VM bytecode.
This design enables end-to-end protection of exception handling while maintaining full compatibility with the unmodified OS runtime.

In addition, \ourproj{} supports 385 x86 instruction encodings and 155 VM handler templates. 
It is implemented using a loosely coupled architecture and an efficient direct-threaded code organization~\cite{piumarta1998optimizing}. 
\ourproj{} provides full support for C++-specific exception-handling semantics and is designed as a reusable research testbed for the academic community.

To evaluate the effectiveness and efficiency of \ourproj{}, we conducted extensive experiments along three dimensions: correctness, resilience, and performance. 
Our results demonstrate that \ourproj{} preserves semantic equivalence with the x86 instruction set architecture under both concrete execution on physical CPUs and symbolic execution. 
In addition, \ourproj{} robustly supports complex control-flow and exception-handling behaviors, enabling correct execution of randomly generated C++ programs.
Furthermore, the \textbf{ABI-Compliant EH Shadowing} mechanism effectively eliminates static fingerprints associated with exception-handling metadata, thereby preventing the reconstruction of exception logic by state-of-the-art reverse-engineering tools such as IDA Pro. 
At the same time, it introduces negligible runtime overhead along the normal execution path and incurs only a minimal increase in binary size.
In summary, we make the following contributions.

\begin{itemize}

\item{
We design and implement \ourproj{}, a comprehensive VM-based obfuscation framework that supports 385 x86 instruction encodings and incorporates state-of-the-art binary protection mechanisms. 
\ourproj{} adopts a loosely coupled architecture that facilitates extensibility and secondary development. 
To the best of our knowledge, \ourproj{} represents the first open-source implementation of a comprehensive VM-based obfuscator. 
}

\item{
We propose a novel exception-aware protection mechanism, termed \textbf{ABI-Compliant EH Shadowing}, in \ourproj{}. This mechanism achieves end-to-end protection of exception handling and effectively narrows the attack surface of the protected binary.
}

\item{
We conduct extensive experiments to evaluate the correctness and resilience of \ourproj{}. 
Our results demonstrate that \ourproj{} supports complex C++ programs, verified against a cumulative codebase exceeding 670K LoC, without leaking exception-handling metadata, while incurring only minimal code-size overhead and negligible runtime overhead.
}

\end{itemize}

The remainder of this paper is organized as follows.
Section~\ref{sec:moti} provides background information and outlines the motivation behind this work.
Section~\ref{sec:method} and Section~\ref{sec:impl} detail the design and the implementation of the \ourproj{} framework.
Section~\ref{sec:evaluation} presents experimental results and provides a comprehensive analysis of each component within \ourproj{}.
Section~\ref{sec:related} reviews related work.
Section~\ref{sec:conclusion} concludes the paper and discusses potential future directions.

\section{Background and Motivation}
\label{sec:moti}

\subsection{Virtualization Obfuscation}
\label{sec:vmtools}

\begin{figure*}[t]
    \centering
    \includegraphics[width=0.9\linewidth]{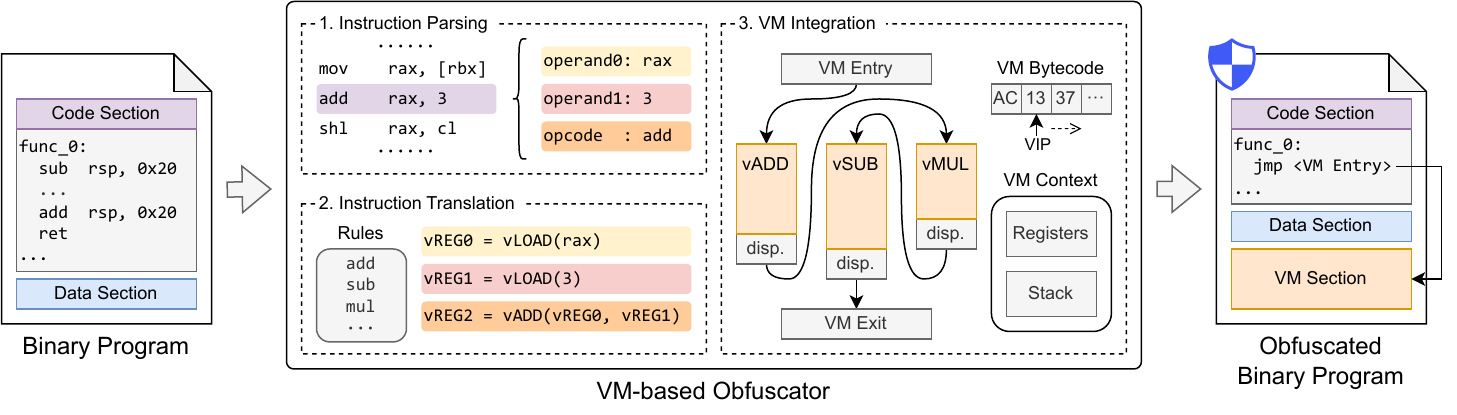}
    \caption{
Workflow of VM-based obfuscation.
The process transforms native instructions through three stages: \emph{instruction parsing}, \emph{instruction translation}, and \emph{VM integration}.
The original code is replaced by a VM entry stub that redirects control flow to the VM interpreter for VM bytecode execution.
\textit{(VIP = virtual instruction pointer; disp. = dispatcher)}
}
    \label{fig:VM}
\end{figure*}

Virtual machine (VM)-based binary obfuscation is a code protection mechanism widely adopted in industry. 
In general, a VM-based obfuscator takes an unprotected binary program as input and produces an obfuscated program represented in VM bytecode. 
As illustrated in Figure~\ref{fig:VM}, the obfuscation workflow consists of three main stages: \emph{instruction parsing}, \emph{instruction translation}, and \emph{VM integration}.
The process begins by identifying the binary code of the target functions to be obfuscated. 
In the \emph{instruction parsing} stage, the obfuscator sequentially parses all native instructions within these functions. 
Subsequently, during the \emph{translation} stage, 
each native instruction is translated into a sequence of micro-operations, referred to as virtual handlers, based on a set of custom mapping rules~\cite{blazytko2017syntia}.
Specifically, these rules define how to decompose complex native instructions into finer-grained primitive operations. For instance, a single native instruction involving memory operands is typically mapped to a sequence of virtual handlers responsible for data loading, calculation, and memory storage.
These micro-operations are encoded as VM bytecode, which serves as the instruction stream to dispatch and execute the corresponding virtual handlers.
Through this translation, program semantics are decoupled from the underlying hardware instruction set.

To hinder analysis-based attacks, sophisticated protection mechanisms are applied to the encoding of these handlers. 
Techniques such as data encoding~\cite{zhou2007information} and instruction metamorphism~\cite{borello2008code} are commonly employed to obscure the actual program behavior. 
Furthermore, obfuscators typically minimize the number of distinct handlers embedded into an obfuscated binary, preventing analysts from reconstructing the complete virtual instruction set from a single sample~\cite{li2022chosen}. 
As a result, RISC-like micro-operation selection strategies are often adopted to reduce the handler set required by the target program.
When designing \ourproj{}, we employed a custom-designed domain-specific language (DSL) for handler generation. This compilation-based approach inherently simplifies the implementation of diverse polymorphic mutations. 
Furthermore, our obfuscator is designed to embed only the minimal set of handlers required by the target program.

Finally, in the \emph{VM integration} stage, the obfuscator assembles these components into the obfuscated binary. It embeds the VM bytecode and the VM interpreter, which comprises the VM entry for context initialization, specific virtual handlers (e.g., vADD, vMUL) with embedded dispatchers, and the VM exit for context restoration. Additionally, a VM context structure is defined to simulate virtual registers and stack. The process concludes by replacing the original function body with a stub that redirects control flow to the VM entry~\cite{li2022chosen}.
At runtime, invoking an obfuscated function transfers control to the VM interpreter, which then parses the encoded bytecode and executes the corresponding handlers sequentially to emulate the original program semantics. 
This transformation fundamentally reconstructs the program’s execution model~\cite{rolles2009unpacking, banescu2016code}.
To further conceal control flow from static analysis, the VM bytecode is often randomized and encrypted, requiring dynamic decryption during execution~\cite{rolles2009unpacking}. 
After the obfuscated computation completes, the VM exits and returns control to the original execution flow.
When designing \ourproj{}, we adopted a pass-driven architecture for bytecode construction. This design decouples the integration logic from specific obfuscation algorithms, allowing researchers to seamlessly plug in custom randomization or encryption schemes as independent passes without modifying the core engine.

Regarding the virtual handler dispatch mechanism, traditional virtualization techniques relied on a centralized dispatcher (e.g., a switch-case structure). However, this design was abandoned due to its conspicuous patterns that facilitate reverse engineering~\cite{sharif2009automatic}. Modern protectors adopt a direct-threaded code structure~\cite{piumarta1998optimizing}. In this design, the dispatcher is embedded at the tail of each handler to calculate and jump to the next address, creating a complex, chained execution path that significantly disrupts the program's control flow~\cite{li2022chosen}.

\subsection{Exception Handling in Operating System}
\label{sec:exceptionhandling}

In modern software systems, exception handling (EH) is a fundamental mechanism for structured error handling and control transfer. 
It is widely used to handle exceptional conditions such as memory exhaustion (e.g., failure of the \texttt{new} operator), invalid container access (e.g., \texttt{std::vector} out-of-bounds access), or network disconnection~\cite{bradley2019study}.
EH is a core language feature in many mainstream programming languages, including Python~\cite{LangPython}, C++~\cite{LangCpp}, and Java~\cite{LangJava}.
At the binary level, specifically for native compiled languages like C++, exception handling behavior is tightly governed by the Application Binary Interface (ABI), which defines how exceptions are propagated, how the call stack is unwound, and how language-specific cleanup routines are invoked~\cite{itanium_abi}. 
Although exceptions may originate from language runtimes or libraries, their propagation relies on ABI-specified conventions that are enforced by the operating system and runtime environment (e.g., glibc or ntdll).

When an exception is raised during function execution, control is transferred to the runtime’s exception dispatcher, which initiates stack unwinding from the point where the exception occurs. 
Stack unwinding involves traversing the call stack frame by frame, restoring register state and invoking language-specific destructors or cleanup handlers as required. 
Correct stack manipulation during this process is essential to preserve program semantics and ensure safe recovery.

Early ABI designs relied on a dedicated frame pointer to identify stack frame boundaries and facilitate unwinding. 
However, to enable more aggressive compiler optimizations such as frame pointer omission and flexible stack layouts, modern ABIs have largely abandoned this approach~\cite{lu2018system, ms_x64_exception}. 
Instead, contemporary exception handling relies on compiler-generated, function-specific metadata that precisely describes how each stack frame should be unwound.
This metadata-driven mechanism allows the runtime to determine, based on the exception’s program counter, how to restore the execution context and propagate the exception correctly. 
The metadata is typically stored in dedicated unwind tables generated at compile time and is directly consulted by the runtime during exception propagation~\cite{dwarf_standard}.

\begin{figure}[t]
    \centering
    \includegraphics[width=0.80\linewidth]{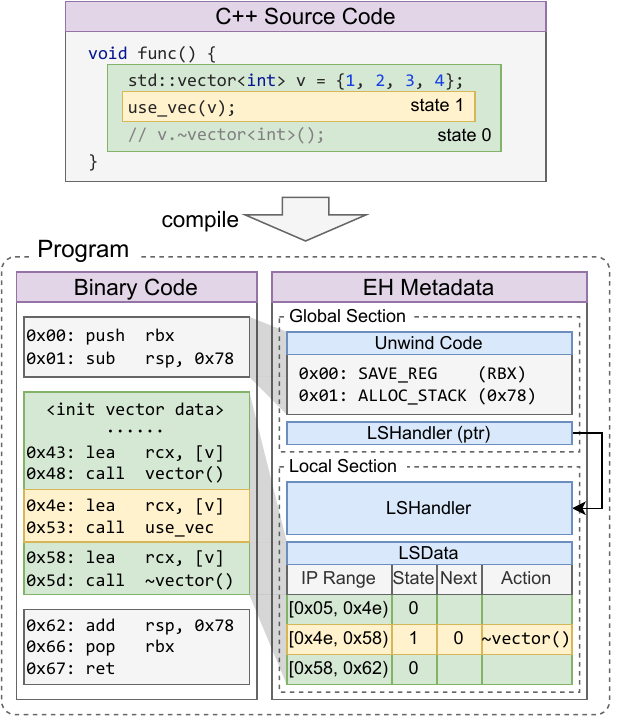}
    \caption{
The intrinsic coupling between binary code and EH metadata.
The \emph{Global Section} describes the static stack frame structure via unwind codes, while the \emph{Local Section} encodes dynamic object lifecycles via LSData.  
This mechanism renders the metadata a faithful reflection of the original program logic.
    }
    \label{fig:mirror}
\end{figure}

The first phase of exception handling is driven by the language runtime in cooperation with the operating system and is commonly referred to as \textbf{Global Unwind}.
Its purpose is to restore the caller’s stack frame and register state while propagating the exception up the call stack. 
To support this process, compilers generate unwind tables that contain a sequence of specialized unwind instructions. 
These instructions describe how each function’s stack frame is constructed in its prologue.
For example, if a function allocates 0x78 bytes of stack space for local variables in its prologue, the corresponding unwind table will contain an instruction such as \texttt{ALLOC\_STACK(0x78)} as shown in Figure~\ref{fig:mirror}. 
When an exception occurs, the runtime interpreter consults these unwind instructions to precisely reverse the effects of the current stack frame, i.e., restoring registers and adjusting the stack pointer, thereby safely unwinding to the caller’s frame.

The second phase of exception handling addresses language-specific cleanup semantics, such as destroying objects that were successfully constructed within a stack frame, and is commonly referred to as \textbf{Local Unwind}. 
This phase is inherently complex and depends on both the programming language and the program implementation.
During local unwinding, the runtime transfers control to a \emph{language-specific handler} (LSHandler), as indicated by a function pointer stored in the exception-handling metadata. 
The LSHandler performs cleanup actions according to compiler-generated \emph{language-specific data} (LSData), which encodes the required destructors and cleanup routines for the program. 
For example, as illustrated in Figure~\ref{fig:mirror}, the LSData may contain a pointer to the destructor \texttt{\textasciitilde{}vector()}.
This process can be highly intricate. 
In languages such as C++, the LSHandler typically implements a finite-state machine (FSM) to ensure that each local object is destructed exactly once and in the correct order, even in the presence of partially constructed stack frames and nested exception scopes.

Figure~\ref{fig:mirror} illustrates an example of exception-handling (EH) metadata. 
The static stack frame information used during the global unwinding phase is denoted as the \emph{Global Section}, while the code and data structures involved in the local unwinding phase are denoted as the \emph{Local Section}.
The \emph{Global Section} describes the per-function stack frame layout used for global unwinding, whereas the \emph{Local Section} encodes intra-frame semantics required for local unwinding.
As shown in the figure, EH metadata encodes a substantial amount of low-level implementation detail, including stack layout, register-saving conventions, and object lifetime information. 
Such information can be exploited to recover structural properties of the program and has been shown to facilitate software reverse engineering~\cite{pang2021towards}.
However, concealing or modifying EH metadata is inherently challenging. 
The unwinding process is strictly defined by the operating system and language runtime in accordance with ABI specifications, and is directly invoked during exception propagation. 
As a result, this mechanism operates largely outside the control of application developers, who must preserve ABI compliance to maintain functional correctness.

\subsection{Exception Handling in Obfuscated Code}
\label{sec:ehinobfuscatedcode}

Preserving semantic consistency is a fundamental requirement for any practical code obfuscation technique. 
Beyond maintaining the functional behavior of the protected binary code along normal execution paths, an obfuscator must also faithfully preserve the semantics of exception handling. 
Exceptions constitute an integral part of program logic in modern software systems, governing error propagation, resource cleanup, and control-flow recovery.
Any discrepancy in exception-handling behavior, such as incorrect stack unwinding, missing cleanup routines, or altered control-flow semantics, can lead to observable misbehavior, compromise program correctness, or introduce new vulnerabilities.
Therefore, effective obfuscation must ensure semantic equivalence across both regular execution and exceptional control paths, while remaining fully compliant with language runtimes and ABI specifications.

In general, code obfuscation protects target code by transforming or replacing native instructions with encoded or encrypted representations. 
In contrast, exception-handling (EH) metadata cannot be directly encrypted, because exception propagation and stack unwinding are governed by the operating system and language runtime, which require EH metadata to strictly conform to the ABI. 
Consequently, any attempt to replace EH metadata with opaque or encrypted data would violate this ABI contract, leading to incorrect stack unwinding and, ultimately, application crashes~\cite{panchenko2019bolt}.

State-of-the-art industrial VM-based obfuscators (e.g., VMProtect and Code Virtualizer) leave exception-handling (EH) metadata unprotected in order to preserve ABI compatibility. 
Specifically, they retain the original EH metadata without modification, including both the global section and local section shown in Figure~\ref{fig:mirror}.
When an exception occurs, execution exits the virtual machine and control is transferred to the native exception-handling routine provided by the operating system, i.e., within the unprotected execution environment. 
The OS and language runtime then perform stack unwinding and invoke the native language-specific handler (LSHandler), as described in Section~\ref{sec:exceptionhandling}.
While this design is simple and robust with respect to correctness, it introduces a protection gap: although the program’s executable code is shielded by virtualization obfuscation, the EH metadata remains fully exposed.

Exposing exception-handling (EH) metadata is particularly risky because most existing attacks against VM-based obfuscation predominantly rely on dynamic analysis~\cite{sharif2009automatic, coogan2011deobfuscation, yadegari2015generic}. 
Attackers typically begin by collecting program execution traces~\cite{yadegari2015generic}. 
To extract meaningful information from these voluminous traces, prior work applies techniques such as simplification and clustering~\cite{coogan2011deobfuscation}, taint analysis~\cite{ming2015loop}, symbolic execution~\cite{banescu2016code}, and program synthesis~\cite{blazytko2017syntia}, with the goal of reconstructing the semantics of the original program.
EH metadata exposes deterministic structural information generated by the compiler, which substantially facilitates this reconstruction process. 

Advanced decompilers, such as IDA Pro~\cite{IDAPro9EHparsing} and Ghidra~\cite{GhidraCode}, now employ sophisticated analysis to reconstruct program logic from EH metadata. Specifically, these tools identify the specific language runtime by matching the LSHandler against known signatures. Based on this identification, they parse the LSData using the corresponding structures. This enables them to recover the exception control flow and merge it with the normal control flow, displaying the combined logic in the decompiled code.

When designing \ourproj{}, we recognize the critical role of exception handling (EH) metadata and provide comprehensive protection for both the binary code and the exception-handling routines, while strictly preserving ABI compliance.

\section{Design of \ourproj{}}
\label{sec:method}

\subsection{System Overview}

\begin{figure}[t]
    \centering
    \includegraphics[width=\linewidth]{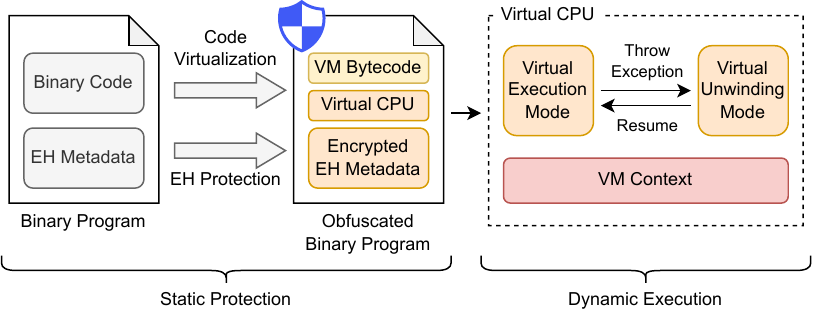}
    \caption{
Overview of \ourproj{}’s exception-aware protection workflow. 
During static transformation, \ourproj{} provides comprehensive protection for both executable code and exception-handling metadata. 
At runtime, protected code executes within \ourproj{}’s VM interpreter. 
After a dummy global unwinding step to preserve ABI compatibility, the remaining exception-handling logic is securely processed within the VM interpreter, achieving both compatibility and strong protection.
    } 
    \label{fig:designoverview}
    
\end{figure}

\ourproj{} is a binary-level, VM-based obfuscator. 
As illustrated in Figure~\ref{fig:designoverview}, \ourproj{} provides comprehensive protection for both the target binary code and the exception-handling procedure.
For executable code, \ourproj{} translates the original instruction sequence into combinations of virtual handlers according to predefined mapping rules, and then generates encrypted VM bytecode, following a workflow similar to existing VM-based obfuscators described in Section~\ref{sec:vmtools}.
In parallel, the exception-handling (EH) metadata is parsed and replaced with a dummy unwind section, while the original metadata is obfuscated and encrypted. 
Finally, all protected components are reassembled together with the required runtime modules—such as the VM interpreter—to produce the fully obfuscated binary.

At runtime, \ourproj{}’s VM interpreter maintains a virtualization environment and takes control of program execution when a protected function is invoked. 
The interpreter executes the obfuscated code by decoding and dispatching VM bytecode, following the workflow described in Section~\ref{sec:vmtools}.
When an exception is raised, control is initially transferred to the operating system runtime. 
However, the dummy unwind section generated by \ourproj{} redirects execution back into a virtual unwinding mode. 
Both normal execution and exception handling are then processed within the same VM context.
This design provides comparable protection strength for both normal execution and exception handling, while ensuring that sensitive exception-related data remains protected.

\subsection{Static Protection for \ourproj{}}

\begin{figure*}[t]
    \centering
    \includegraphics[width=\linewidth]{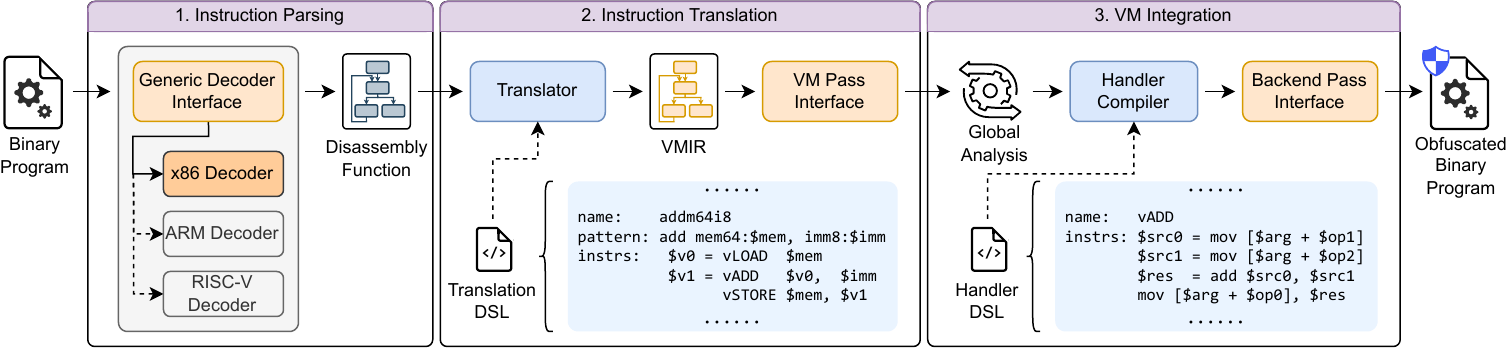}
    \caption{
The architecture of \ourproj{}’s static protection engine. The workflow operates in three decoupled stages: \emph{Instruction Parsing}, \emph{Instruction Translation}, and \emph{VM Integration}. \ourproj{} employs custom domain-specific languages (translation and handler DSL) to decouple obfuscation logic from the underlying architecture. Furthermore, it adopts a pass-driven design with dedicated interfaces (VM and backend pass) to facilitate flexible transformation.
    }
    \label{fig:extensibility}
\end{figure*}

Similar to previously proposed VM-based obfuscation tools, \ourproj{} follows a three-phase design consisting of \emph{Instruction Parsing}, \emph{Instruction Translation}, and \emph{VM Integration} as mentioned in Section~\ref{sec:vmtools}. 
However, \ourproj{} adopts a modular, loosely coupled architecture that allows researchers to extend or replace individual components with minimal effort.
To reduce inter-stage dependencies, we employ a pass-driven architecture, similar to that used in widely deployed compiler frameworks such as LLVM. 
In addition, \ourproj{} defines a domain-specific language (DSL) to facilitate cross-architecture instruction translation and VM handler generation.

\textbf{Instruction Parsing} stage transforms raw binary code into a structured intermediate representation. 
Specifically, it decodes native instructions from the target functions and reconstructs them into instruction basic blocks and control-flow graphs (CFGs).
\ourproj{} adopts an object-oriented design that encapsulates the decoding logic of a specific instruction set architecture (ISA) behind a generic interface. 
Following the design philosophy of LLVM and MLIR, we construct basic blocks and CFGs using a unified representation. 
This abstraction enables the decoder component to be easily replaced, allowing the framework to be adapted to other ISAs (e.g., ARM or RISC-V) without modifying the core analysis logic.

In general, it is theoretically impossible to perfectly distinguish code from data in arbitrary binary programs. 
This limitation is rooted in the Halting Problem of Turing Machine. 
Consequently, existing binary analysis tools necessarily rely on conservative assumptions or heuristics~\cite{schwartz2010all}.
To address this challenge in practice, \ourproj{} implements a recursive descent disassembly algorithm, a standard approach whose correctness has been widely validated in prior binary analysis works~\cite{schwarz2002disassembly, andriesse2016depth}.
Instead of linearly sweeping the entire code section, this algorithm starts from the target function entry points and recursively traverses the control flow graph (CFG) by following linear execution paths and static branch targets.
This approach provides conditional correctness: it accurately distinguishes instructions from data under the condition that the code is statically reachable through explicit control flow.
Although this is an approximation algorithm that may miss code hidden behind complex indirect jumps, it effectively minimizes false positives compared to linear sweep methods.
For the sequential decoding within each identified basic block, we employ Zydis~\cite{zydis}.
Zydis is a mature and verified disassembly library that reliably parses contiguous variable-length x86 instructions, ensuring parsing correctness at the instruction level.

Building upon this instruction-level decoding, we implement a robust reconstruction engine to recover the function-level CFG.
This engine comprehensively handles standard control transfers, including conditional branches (\texttt{jcc}), intra-procedural direct jumps (\texttt{jmp}), and function returns (\texttt{ret}).
Additionally, it correctly identifies software interrupts (\texttt{int3}), which typically serve as execution barriers padding the unreachable code following non-returning function calls.
Beyond these explicit transfers, implicit control flow presents significant challenges.
To address this, \ourproj{} integrates established control flow recovery techniques commonly employed in modern decompilers to resolve two complex scenarios: indirect jumps (specifically \texttt{switch} statements) and tail calls.

First, regarding \texttt{switch} statements, compilers typically translate sparse cases into a cascade of conditional branches (i.e., \texttt{if-else} chains).
However, for performance optimization, dense case values are often compiled into jump tables---arrays of code pointers stored in the data section.
At runtime, the program calculates an index to retrieve the target address from the table and performs an indirect jump.
To recover the complete control flow, \ourproj{} analyzes the memory addressing pattern of the indirect jump instruction to locate the jump table.
It then parses the pointers stored within the table to identify all potential successor basic blocks and adds them to the recursive traversal queue.

Second, \ourproj{} addresses the ambiguity introduced by Tail Call Optimization (TCO).
TCO is a compiler optimization that transforms a function call immediately followed by a return (i.e., \texttt{call} then \texttt{ret}) into a single direct jump instruction (\texttt{jmp}), thereby reusing the current stack frame.
In binary analysis, this creates a specific challenge: a \texttt{jmp} opcode may represent either an intra-procedural control transfer (a local jump) or an inter-procedural tail call.
To distinguish between these scenarios, \ourproj{} leverages the Exception Handling (EH) metadata.
We query the metadata table to retrieve the precise address range $[start\_pc, end\_pc)$ of the current function.
We then evaluate the target address of the \texttt{jmp} instruction: if the target falls within this range, it is classified as a local jump between basic blocks; otherwise, it is identified as a tail call, marking the termination of the current function's control flow.

\textbf{Instruction Translation} stage transforms target functions from native instructions into VM-specific instructions. 
In \ourproj{}, we define a RISC-like instruction set that serves as a high-level intermediate representation (IR), referred to as \textbf{VMIR}. 
Translation is performed using a rule-based mechanism: for each native instruction, a predefined sequence of VMIR instructions is specified to preserve equivalent semantics. 
During translation, \ourproj{} matches each native instruction against these rules and emits the corresponding VMIR sequence, as illustrated in Figure~\ref{fig:extensibility}.

To improve scalability and extensibility, we introduce a domain-specific language for specifying translation rules (referred to as \emph{Translation DSL} in Figure~\ref{fig:extensibility}), inspired by the design of LLVM TableGen~\cite{lattner2004llvm}. 
As shown in the second stage of Figure~\ref{fig:extensibility}, the DSL describes how a native instruction (e.g., \texttt{addm64i8}) is translated into a sequence of VMIR instructions. 
Using this DSL, \ourproj{} currently supports 385 x86 instruction encodings spanning 42 opcode types, which is sufficient to handle general-purpose applications, as verified by our extensive evaluation using the cryptographic algorithms and 1,000 randomized C++ programs generated by YARPGen~\cite{livinskii2020random}.
Moreover, the DSL is decoupled from the translation engine, allowing new instructions to be added with minimal effort.

It is worth mentioning that \ourproj{} exposes an interface for pass-based transformation, following the design principles of modern compiler frameworks. 
VMIR is represented as a mutable graph, enabling developers to apply additional semantic obfuscation or optimization passes before lowering VMIR into VM bytecode.

\textbf{VM Integration} stage produces the final obfuscated binary by generating a customized VM interpreter together with encrypted VMIR bytecode. 
\ourproj{} performs a global analysis over the translated VMIR to identify the minimal set of VM-specific instructions required by the target program. 
For each such instruction, a corresponding handler is integrated into the VM interpreter. 
Each handler consists of a sequence of native instructions that implements semantics equivalent to the associated VM-specific instruction.
During integration, handlers are dynamically compiled into native assembly code. 
This compilation-based approach enables greater diversity in handler implementations, as discussed in Section~\ref{sec:vmtools}, thereby increasing resistance to analysis-based attacks.

To simplify handler implementation across different architectures, we further define a handler description language, referred to as \textbf{Handler DSL}. 
This DSL is decoupled from the handler generation engine and can be readily ported across architectures. 
Crucially, the VM instruction set defined in \ourproj{} encompasses a fundamental subset of operations covering arithmetic logic (e.g., \texttt{ADD}, \texttt{SUB}), memory manipulation (e.g., \texttt{LOAD}, \texttt{STORE}), and control flow transfer (e.g., \texttt{JCC}, \texttt{JMP}).
This combination theoretically guarantees Turing completeness, ensuring that any arbitrary computation can be represented.
For each instruction, \ourproj{} provides corresponding handler specifications in Handler DSL for the x86 architecture.

In addition, \ourproj{} exposes backend interfaces that allow developers to inject further variations into handler code generation through pass-based transformations.

\subsection{Dynamic Execution of \ourproj{}}

The dynamic execution process is initiated by the VM entry stub, which acts as the gateway between the native and virtual worlds. 
Upon invoking a protected function, control is immediately transferred to this stub. 
Its primary responsibility is to initialize the virtualization environment by preserving the current physical CPU state---including general-purpose registers and processor flags (e.g., \texttt{EFLAGS})---into the VM context, which is allocated directly on the stack to support thread safety. 
Subsequently, the virtual instruction pointer (VIP) is set to the entry point of the target function's VM bytecode. 
The VM interpreter then enters its main execution process, operating as a virtual CPU: it cyclically fetches the next opcode pointed to by the VIP, decodes the corresponding virtual handler address and operands, and executes the handler to emulate the semantics of the original instruction. 
This process continues until the virtual execution concludes. 
Finally, the VM exit routine restores the physical CPU state from the VM context and transfers control back to the native execution flow, ensuring a seamless transition for the calling function.

Finally, the generated bytecode is organized using the direct-threaded code structure, where the dispatch logic for the next instruction is embedded at the tail of each handler. This design discards the centralized dispatcher in traditional virtual CPU, which is prone to pattern matching. By decentralizing the dispatch control flow, it increases the topological complexity of the control flow graph, thereby enhancing resilience against adversarial analysis.

\subsection{Exception Handling Procedure Protection}

\begin{figure}[t]
    \centering
    \includegraphics[width=\linewidth]{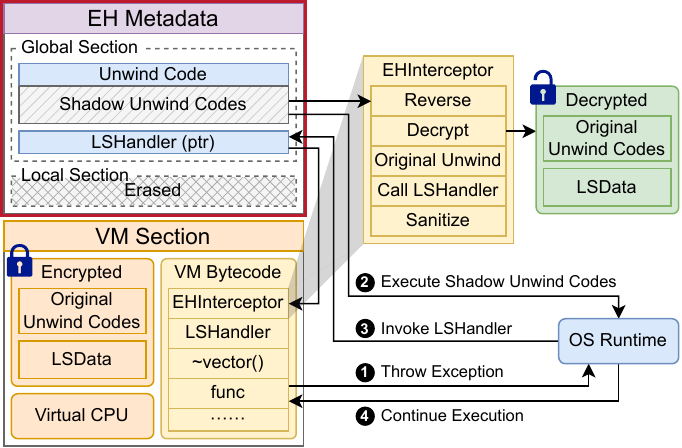}
    \caption{
Overview of the runtime mechanism of ABI-Compliant EH Shadowing in \ourproj{}.
The detailed structure of the EH metadata within the red box corresponds to the layout depicted in Figure~\ref{fig:mirror}.
    }
    \label{fig:EHProtect}
\end{figure}

The \textbf{ABI-Compliant EH Shadowing} mechanism is illustrated in Figure~\ref{fig:EHProtect}. 
During static protection, \ourproj{} transforms the program's EH metadata into an ABI-compliant shadow form by introducing \emph{shadow unwind codes} and redirecting the LSHandler pointer to a shadow handler.
At runtime, this shadow handler redirects control flow to the protected exception handler inside the virtual machine, where the logic is obfuscated and securely executed.

As shown on the left side of Figure~\ref{fig:EHProtect}, the \ourproj{} engine removes the original unwind codes in the \emph{Global Section} and replaces them with a carefully constructed set of \emph{shadow unwind codes}.
As the global unwinding phase is driven by the OS runtime, these shadow unwind codes strictly conform to the ABI, ensuring that the OS can safely execute the unwinding procedure and reach the language-specific handler dispatch. 
Importantly, the \emph{shadow unwind codes} are not chosen arbitrarily: \ourproj{} restricts them to ABI-valid operations that do not clobber architectural states potentially consumed by the subsequent local unwinding logic. 
In particular, we avoid unwind operations that restore important non-volatile registers from the stack (e.g., \texttt{SAVE\_REG(RBP)}), because overwriting the frame pointer with stack-resident values can invalidate later stack traversal and handler execution. 
Under these constraints, the shadow unwind codes remain semantically decoupled from the original exception-handling metadata and act only as an ABI-valid placeholder that prevents structural leakage.
These temporary effects are subsequently rolled back by \texttt{EHInterceptor}, which restores the correct unwinding state inside the protected VM context.

For the \emph{Local Unwind} phase, \ourproj{} erases the \emph{Local Section} and relocates the original language-specific handler (LSHandler) and its associated code fragments (such as destructors and catch blocks) into the virtual machine, where they are translated into obfuscated VM bytecode. 
In parallel, the original unwind codes and language-specific data (LSData) are encrypted and encapsulated within the virtual machine.

To coordinate exception handling at runtime, \ourproj{} implements a VM-resident helper function, \texttt{EHInterceptor}. 
This function first reverses the stack effects introduced by the shadow unwind codes, 
then decrypts the original global codes and LSData, and finally performs correct exception propagation and stack unwinding based on the recovered information. 
After completing the global unwinding phase, \texttt{EHInterceptor} transfers control to the relocated LSHandler inside the VM and sanitizes the decrypted data before returning.
The VM entry stub of \texttt{EHInterceptor} is installed as the LSHandler pointer in the \emph{Global Section}, fully conforming to ABI requirements.

The right side of Figure~\ref{fig:EHProtect} illustrates the dynamic execution of the \textbf{ABI-Compliant EH Shadowing} mechanism. 
When the program throws an exception \blackcircled{1}, control flow is first transferred to the operating system runtime. 
The OS reads and executes the shadow unwind codes outside the virtual machine and temporarily modifies the stack frame \blackcircled{2}.
Subsequently, when the OS attempts to invoke the language-specific handler (LSHandler) via the LSHandler pointer, control flow is redirected to the VM-resident interceptor function, \texttt{EHInterceptor} \blackcircled{3}. Inside \texttt{EHInterceptor}, \ourproj{} rolls back the stack pointer to undo the effects of the shadow unwinding, and then performs the correct exception propagation and stack unwinding within the protected, obfuscated execution context. 
Once the exception is successfully handled, normal program execution resumes \blackcircled{4}.

\section{Implementation}
\label{sec:impl}

\ourproj{} is implemented from scratch in C++. 
The core framework consists of 9,415 lines of code. 
In addition, 5,471 lines of domain-specific language (DSL) code are used to implement support for the x86 instruction set. 
The current prototype supports 385 instruction encoding formats under the x86 architecture and integrates 155 built-in VM handler templates.

To ensure comprehensive instruction set compatibility, \ourproj{} employs a context-switching fallback strategy for instructions that are not fully covered by the instruction translation module (e.g., obscure system instructions or memory-access instructions with a \texttt{LOCK} prefix). 
Specifically, the system temporarily exits virtualization mode, restores the native execution context to execute the single instruction directly, and then immediately re-enters virtualization to reclaim control of the execution flow. 
This hybrid execution mechanism guarantees functional correctness for arbitrary instruction streams and is also widely adopted by industrial-grade obfuscators~\cite{li2022chosen}.

To support multi-threaded applications, \ourproj{} follows lock-free design principles by allocating the VM context directly on the runtime stack of the executing thread. 
Because stack memory is inherently thread-local, this design naturally provides isolated VM contexts for concurrent threads, enabling stable support for multi-threaded programs without introducing additional synchronization overhead.

\section{Evaluation}
\label{sec:evaluation}

\begin{table}[t]
\centering
\caption{Feature comparison between binary-level VM-based obfuscators.}
\label{tab:comparison}
\resizebox{\linewidth}{!}{%
\begin{tabular}{llccc}
\toprule
\makecell{Obfuscator} &
\makecell{Dispatch Mechanism} & 
\makecell{EH\\Support} & 
\makecell{EH Metadata\\Protection} 
\\
\midrule
Obsidium         & Decode-Dispatch Loop    & \ding{55} & --        \\
VMProtect (v2)   & Decode-Dispatch Loop    & \ding{51} & \ding{55} \\
VMProtect (v3)   & Direct-Threaded Code    & \ding{51} & \ding{55} \\
Code Virtualizer & Direct-Threaded Code    & \ding{51} & \ding{55} \\
\midrule
\ourproj{}       & Direct-Threaded Code    & \ding{51} & \ding{51} \\
\bottomrule
\end{tabular}%
}
\end{table}

Table~\ref{tab:comparison} compares the features of representative binary-level VM-based obfuscators. 
Existing designs largely fall into two execution models: earlier obfuscators often adopt a centralized decode-dispatch loop, while more recent obfuscators favor direct-threaded code to reduce recognizable dispatch patterns and increase control-flow complexity. 
Regarding EH, existing obfuscators either do not support EH semantics or leave EH metadata readily accessible to preserve ABI compatibility, thereby exposing structural information. 
In contrast, \ourproj{} uses a direct-threaded interpreter with EH support and protects EH metadata.

To evaluate the effectiveness of \ourproj{}, we conduct a comprehensive set of experiments. 
We first assess semantic correctness to verify that program behavior is preserved under obfuscation. 
We then measure both the static file-size overhead introduced during static protection and the runtime overhead incurred during dynamic execution.
In particular, for the \textbf{ABI-Compliant EH Shadowing} mechanism, we evaluate the resilience of exception-handling metadata protection as well as the additional overhead introduced by exception shadowing.

For comparison, we implement two variants of \ourproj{}.
(1) \ourproj{}-Base, which enables only code obfuscation while retaining the original exception-handling mechanism, thereby emulating the behavior of existing VM-based obfuscators without EH protection; and
(2) \ourproj{}-EHProtect, which enables both code obfuscation and the ABI-Compliant EH Shadowing mechanism.
In addition, we include two industry-grade obfuscation tools—VMProtect and Code Virtualizer for performance comparison.
All experiments were conducted on a desktop system equipped with an Intel Core i9-13900K CPU and 64 GB of RAM, running 64-bit Windows 11. 
We select a diverse set of source programs for evaluation to assess both the correctness and performance of binary obfuscation and exception handling. 
All binaries are compiled using MSVC v19.44.
This setup reflects the execution environment targeted by the obfuscated binaries.

\subsection{Correctness}

We have evaluated the correctness of \ourproj{} across three dimensions: instruction-level semantics, program-level semantics, and the exception handling mechanism.

\textbf{Instruction-Level Semantic Validation.} 
We first evaluate the correctness of instruction translation and emulation. 
Because the majority of obfuscation is performed on the x86 instruction set architecture, we focus our evaluation on x86 instructions, which are sufficiently complex to stress the translation and execution pipeline. 
Correctness is assessed by comparing execution results between native and obfuscated code. 
To ensure comprehensive coverage, we developed an automated test case generator based on AsmJit~\cite{asmjit}.
By iterating through the x86 opcode table and recursively combining diverse operand types---including general-purpose registers, immediates with boundary values (e.g., \texttt{0x00}, \texttt{0xFF}, \texttt{0x80000000}), and complex memory addressing modes (e.g., \texttt{[rbx+rcx*8+0x1000]})---the generator produced 1,120,279 assembly-level test cases covering 413 x86 opcodes, more than the 42 opcodes supported by \ourproj{}.
This discrepancy is intentional: it allows us to simultaneously verify the correctness of the translation rules for supported instructions and the reliability of the context-switching fallback mechanism for unsupported ones.
Specifically, for each test case, we initialize both the native instruction sequence and the corresponding obfuscated code generated by \ourproj{} with identical CPU contexts, and then compare the post-execution states, including general-purpose registers and the \texttt{EFLAGS} register. 
To precisely control the execution environment, the harness utilizes OS-level context switching primitives to inject an identical architectural state (including general-purpose registers and \texttt{EFLAGS}) before execution and capture the post-execution state. 
A global exception handler is registered to intercept and filter out test cases that trigger invalid memory accesses due to randomized memory operands. 
To exclude invalid memory operations, we apply a strict filtering criterion: a test case is considered valid only if the native instruction executes successfully without triggering an exception. 
Across all valid test cases, \ourproj{} produces execution states that are fully consistent with native execution.

To more exhaustively explore the instruction execution state space, we further conduct symbolic execution-based verification using angr~\cite{shoshitaishvili2016sok}.
We constructed a fully symbolic execution environment where 15 general-purpose registers (from \texttt{RAX} to \texttt{R15}, excluding \texttt{RSP}) and the \texttt{EFLAGS} register are initialized as unconstrained symbolic variables. 
We also create a symbolic memory region which is pre-filled with unconstrained symbolic values, allowing us to verify the correctness of value propagation during memory reads and writes.
We executed both the native and obfuscated instructions within this environment and compared the final symbolic expressions of the registers and the dedicated memory region. 
Note that instructions explicitly manipulating the stack or control flow (e.g., \texttt{PUSH}, \texttt{POP}, \texttt{CALL}) are inherently coupled with the program-level semantic and cannot be verified at the instruction level; therefore, their correctness is rigorously evaluated in the subsequent Program-Level Semantic Validation.
This approach allows us to validate memory-access instructions that were excluded from dynamic testing and makes the verification independent of concrete memory layouts. 
We reuse the instruction test cases generated in the previous evaluation, and \ourproj{} successfully passes all symbolic execution checks.

\textbf{Program-Level Semantic Validation.} 
To further evaluate the robustness of \ourproj{} in handling complex control flows, data dependencies, and compiler optimizations, we conducted extensive differential testing using the YARPGen~\cite{livinskii2020random} compiler fuzzer (v2.0). 
Unlike the widely used Csmith~\cite{yang2011finding}, which is restricted to the C language, YARPGen supports diverse C++ constructs essential for our evaluation.
We generated 1,000 distinct C++ test programs by iterating the generation seed (from 1 to 1,000), resulting in a cumulative code base of 678,038 lines of code.
These generated programs are designed to stress-test compiler optimizers, featuring diverse language constructs such as multi-dimensional arrays, pointer arithmetic, deep nested loops, and mixed bitwise-arithmetic operations.
Each generated test case follows a strict execution lifecycle: it initializes global scalars and arrays with deterministic constants, performs a high-complexity core computation involving intensive state mutations, and finally traverses all global variables to compute a cumulative checksum hash.
We compiled these programs using MSVC with the \texttt{/O2} flag to enable maximum optimization.
We execute both the native binaries and their obfuscated counterparts produced by \ourproj{}, and compare their outputs. 
Across all test cases, the obfuscated programs exhibit behavior identical to their native versions, demonstrating that \ourproj{} correctly preserves semantics for programs with complex control-flow structures.

\textbf{Exception Handling Semantics Validation.}
We construct a dedicated exception-handling (EH) test suite to verify the correctness of the virtual unwinding mechanism, which is not covered by general-purpose benchmarks. 
These tests include 8 hand-crafted test cases implemented in a C++ program, 
which covers two dimensions of exception handling: control flow propagation and object lifetime management.

The first category validates the management of control flow transfers required by the C++ ABI through four C++ scenarios.
(1) We begin by validating inheritance-aware exception matching, where an exception object is thrown and verified to be caught correctly by a \texttt{catch} handler of either its exact type or an ancestor type.
(2) We then evaluate the exception rethrowing mechanism by inserting an empty \texttt{throw} statement inside a \texttt{catch} block, and verify that \ourproj{} correctly propagates the exception to an outer \texttt{catch} handler. 
We further confirm that the rethrown exception refers to the same exception object as the one caught in the inner handler, rather than creating a new exception instance.
(3) Additionally, we evaluate cross-frame unwinding by throwing an exception in a nested virtualized function, and verify that \ourproj{} propagates it across multiple virtualized stack frames until it reaches a matching outer \texttt{catch} handler.
(4) Finally, we evaluate lambda compatibility. 
In C++, a lambda function is an inline anonymous function object that can be used like a regular function. 
We throw an exception from a lambda function invoked within a virtualized function, and verify that \ourproj{} propagates the exception out of the lambda function and that it is handled by an outer \texttt{catch} handler.

The second category evaluates object lifetime management in \ourproj{}, focusing on the correct execution of destructors, including both destruction order and memory reclamation behavior. 
We construct a set of complex scenarios, including: 
the ordering of local object destruction; 
the automatic memory-release behavior of smart pointers such as \texttt{std::unique\_ptr}; 
the destruction of complex objects from the C++ standard library that trigger internal STL logic; 
and composite objects consisting of nested or container-based sub-objects (e.g., \texttt{std::vector} containing user-defined objects).

To verify correctness, we instrument the test suite with a global lifecycle-monitoring mechanism. 
A global atomic counter is used to track object lifetimes: the counter is incremented when an object is constructed and decremented in its corresponding destructor. 
After the exception is handled, we check that the counter returns to zero, confirming that all objects allocated prior to the exception have correctly invoked their destructors.

We also include the LLVM test suite~\cite{llvm-test-suite} in our evaluation. 
From the suite, we extracted 23 test cases related to exception handling. 
We exclude \texttt{spirit}, which is not supported by the MSVC compiler used in our experiments. 
These benchmarks exercise a wide range of corner-case exception-handling behaviors and provide complementary coverage beyond our manually constructed tests.

We execute each test case ten times. 
Across all cases, the protected programs exhibit behavior consistent with their native counterparts. 
Specifically, the programs do not crash unexpectedly, the execution path reaches the designated \texttt{catch} handler, and all allocated memory is correctly reclaimed following exception handling.

\subsection{Overhead of \ourproj{}}

\begin{table*}[t]
\caption{
End-to-end performance and size overhead on the Loki Benchmark.
\textit{Time and size factor are normalized to native binaries.
Values in parentheses indicate the absolute file size increase ($\Delta$ KB) introduced by the EHProtect mechanism.}
}
\label{tab:loki_time}
\resizebox{\linewidth}{!}{%
\begin{tabular}{@{}lrrrrrrrrrr@{}}
\toprule
                                & \multicolumn{5}{c}{Time Factor}                                                                                                  & \multicolumn{5}{c}{Size Factor}                                                                                                  \\
                                & \multicolumn{1}{c}{AES} & \multicolumn{1}{c}{DES} & \multicolumn{1}{c}{MD5} & \multicolumn{1}{c}{RC4} & \multicolumn{1}{c}{SHA1} & \multicolumn{1}{c}{AES} & \multicolumn{1}{c}{DES} & \multicolumn{1}{c}{MD5} & \multicolumn{1}{c}{RC4} & \multicolumn{1}{c}{SHA1} \\ \midrule
VMProtect (Virtualization)      & 7208.88                 & 6696.38                 & 19120.11                & 6805.70                 & 12695.65                 & 180.93                  & 141.88                  & 177.82                  & 184.99                  & 188.27                   \\
VMProtect (Ultra)               & 22046.31                & 23559.08                & 20474.93                & 10240.98                & 36431.06                 & 228.27                  & 203.83                  & 225.74                  & 213.98                  & 234.17                   \\
Code Virtualizer (Fish White)    & 5959.51                 & 7101.85                 & 2485.15                 & 3887.08                 & 10979.13                 & 46.24                   & 48.10                   & 41.67                   & 39.19                   & 48.99                    \\
Code Virtualizer (Dolphin Black) & 8608.42                 & 11648.27                & 3156.17                 & 4034.45                 & 12440.62                 & 68.24                   & 74.99                   & 82.42                   & 77.23                   & 72.96                    \\
Tigress                         & 246.88                  & 59.83                   & 71.39                   & 42.02                   & 85.08                    & 8.75                    & 11.06                   & 8.23                    & 3.11                    & 4.72                     \\
\ourproj{}-Base                    & 259.70                  & 231.87                  & 179.33                  & 120.64                  & 337.23                   & 3.47                    & 10.04                   & 2.82                    & 2.10                    & 3.39                     \\
\ourproj{}-EHProtect               & 260.81                  & 231.05                  & 179.97                  & 120.81                  & 336.61                   & 5.58                    & 11.29                   & 5.04                    & 4.48                    & 5.65                     \\
                                &                         &                         &                         &                         &                          & (+37.24KB)              & (+37.02KB)              & (+37.64KB)              & (+37.39KB)              & (+37.53KB)               \\ \bottomrule
\end{tabular}%
}
\end{table*}

Code obfuscation is typically applied to security-critical components of a program, such as signature verification and cryptographic routines, which are representative targets for protecting intellectual property. 
Accordingly, we select five programs implementing widely used cryptographic algorithms, AES, DES, MD5, RC4, and SHA-1, following the evaluation methodology of~\cite{schloegel2022loki}. 
For each target, we execute the binary 10,000 times with random inputs and record the average execution time.
We additionally include two distinct categories of obfuscators:
We compare \ourproj{} against two types of tools:
(1) Industry-standard binary obfuscators, including VMProtect (v3.9.4) and Code Virtualizer (v3.1.4.0). We evaluate these tools using multiple configurations to serve as the baseline for binary-level virtualization.
(2) Tigress~\cite{tigress} (v4.0.10), a comprehensive source-level obfuscator. 
We executed Tigress within the WSL environment. To align with our virtualization focus, we configured it with the flags \texttt{-{}-Transform=Virtualize} and \texttt{-{}-VirtualizeDispatch=direct}, representing the state-of-the-art in source-based virtualization.

\textbf{Runtime Performance.}
As shown in Table~\ref{tab:loki_time}, the runtime overhead of \ourproj{} ranges from a factor of $120\times$ to $337\times$ compared to the native execution.
While this overhead is inherent to the virtualization-based execution model---where virtual CPU instructions are emulated via a VM interpreter---it represents a significant performance advantage over commercial binary protectors.
Specifically, VMProtect and Code Virtualizer impose a slowdown ranging from $2,485\times$ to $36,431\times$.
This large difference is mainly due to the complexity of the current implementations.
As analyzed in recent studies~\cite{zhang2025inspecting}, commercial obfuscators like VMProtect employ aggressive handler diversification strategies, such as one-to-many opcode mappings, where a single virtual operation is expanded into multiple structurally distinct handlers to evade pattern matching. 
Furthermore, they typically inject intensive arithmetic obfuscation and immediate value decryption logic within each handler to thwart symbolic execution. 
In contrast, \ourproj{} currently uses a simple instruction mapping. We have not yet enabled complex handler expansion or duplication passes.
While \ourproj{} supports the integration of such hardening passes for enhanced security, the baseline configuration prioritizes execution efficiency, resulting in a more lightweight virtualization overhead.
Compared to Tigress, which incurs a runtime overhead ranging from $42\times$ to $247\times$, \ourproj{} exhibits higher overhead.
This is expected because of the difference between source-level and binary-level protection.
Tigress works on source code, so the compiler can still optimize the code after obfuscation (e.g., constant propagation, dead code elimination).
On the other hand, \ourproj{} works on binaries, incurs the unavoidable overhead associated with binary lifting and the reconstruction of execution contexts, which limits the scope of low-level optimizations available to the obfuscator.

\textbf{File Size Overhead.}
For disk size, \ourproj{} increases the file size by $2.1\times$ to $11.2\times$. This growth is comparable to that of Tigress.
This is much more efficient than VMProtect and Code Virtualizer, which increase the size by up to $234\times$.
The huge size increase in commercial tools comes from their handler duplication mechanism~\cite{zhang2025inspecting}, which creates many copies of handlers for the same operation.
In contrast, \ourproj{} currently adopts a minimalistic design philosophy, embedding only the essential set of virtual handlers required by the target binary.
This approach prioritizes functional correctness and execution efficiency in the current prototype.
However, the loosely coupled architecture of \ourproj{} facilitates extensibility; advanced hardening passes, such as handler mutation and duplication, can be integrated in future developments to enhance security at the cost of increased file size.

\subsection{Resilience of EH Shadowing}

\begin{table}[t]
\caption{
Diversity Analysis of Shadow Unwind Codes
}
\label{tab:loki_unique}
\begin{tabular}{@{}crrrrrr@{}}
\toprule
CodeCount & \multicolumn{1}{c}{AES} & \multicolumn{1}{c}{DES} & \multicolumn{1}{c}{MD5} & \multicolumn{1}{c}{RC4} & \multicolumn{1}{c}{SHA1} & \multicolumn{1}{c}{Avg.} \\ \midrule
1           & 12                      & 12                      & 12                      & 12                      & 12                       & 12                       \\
2           & 113                     & 117                     & 108                     & 101                     & 96                       & 107                      \\
3           & 587                     & 635                     & 627                     & 635                     & 638                      & 624                      \\
4           & 970                     & 984                     & 981                     & 978                     & 993                      & 981                      \\
5           & 1000                    & 1000                    & 1000                    & 1000                    & 1000                     & 1000                     \\ \bottomrule
\end{tabular}
\end{table}

We evaluate the effectiveness of the \textbf{ABI-Compliant EH Shadowing} mechanism by examining its ability to eliminate static signatures and disrupt the parsing workflows of standard binary analysis tools. 
To this end, the global unwind section is randomized to prevent information leakage, and we assess both the correctness and diversity of the generated shadow unwind code.

Following the evaluation methodology of~\cite{schloegel2022loki}, we select five cryptographic programs, AES, DES, MD5, RC4, and SHA-1, as workloads. 
For each program, we generate 1,000 obfuscated binaries while varying the length of the shadow unwind code sequence from 1 to 5. 
Table~\ref{tab:loki_unique} reports the number of distinct shadow unwind code sequences observed under each configuration, while the ``CodeCount'' column denotes the sequence length of the shadow unwind codes generated.
During the generation of the shadow unwind section, \ourproj{} admits 12 different valid unwind code types, increasing the sequence length substantially expands the space of possible shadow unwind configurations. 
When the sequence length is set to 5, all 1,000 generated instances are distinct, demonstrating a high degree of diversity. 
Crucially, all generated shadow unwind sequences are syntactically valid and fully compliant with x86 stack-unwinding specifications, making them statically indistinguishable from compiler-generated unwind metadata.

To evaluate resistance against automated static analysis, we analyze the obfuscated binaries using IDA Pro, 
the industry-standard reverse engineering platform known for its advanced decompilation capabilities.
IDA Pro reconstructs high-level exception-handling structures by employing a hierarchical analysis strategy.
First, it inspects the LSHandler pointer referenced in the unwind metadata and performs pattern matching against its internal signature library of known runtime personality routines.
Once the handler type is identified, IDA Pro parses the associated LSData according to the matched format.
Particularly for C++ binaries, this data encodes a complex state machine, defining instruction pointer ranges, state transitions, and the specific cleanup actions (e.g., destructors) required for each state (as shown in Figure~\ref{fig:mirror}).
By resolving these states, IDA Pro merges the exceptional control flow edges with the standard execution flow, synthesizing the complete \texttt{try-catch} logic presented in the decompiled code.

Our experiments show that IDA Pro successfully reconstructs exception-handling structures for binaries protected by VMProtect and \ourproj{}-Base, but fails to do so for \ourproj{}-EHProtect. 
This failure is attributable to the ABI-Compliant EH Shadowing mechanism.
First, IDA Pro relies on the structure of EH metadata and the signature of the language-specific handler (LSHandler) to identify exception-handling constructs. 
By redirecting the LSHandler to the VM-resident \texttt{EHInterceptor}, \ourproj{} breaks the standard runtime patterns expected by the tool, preventing automatic recognition. 
Second, even when the metadata region is manually specified, IDA Pro is unable to parse meaningful data structures because the genuine language-specific data (LSData) is encrypted and encapsulated within the virtual machine.

\subsection{Overhead of EH Shadowing}

\begin{figure*}[t]
    \centering
    \includegraphics[width=\linewidth]{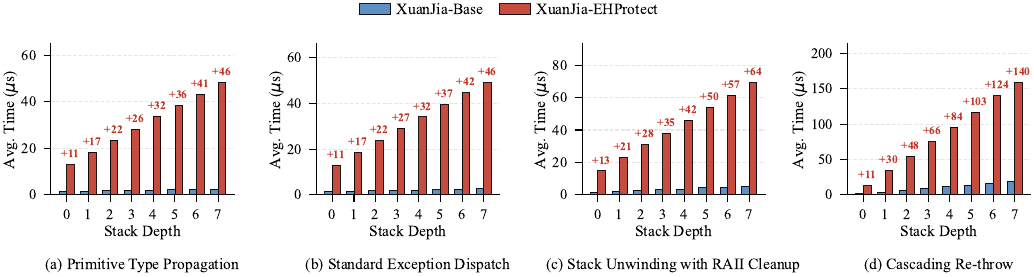}
    \caption{
Runtime overhead of virtual unwinding.
Average execution time of \ourproj{}-Base and \ourproj{}-EHProtect across varying stack depths.
The numeric annotations indicate the relative slowdown factor of \ourproj{}-EHProtect compared to the baseline.
    }
    \label{fig:result_eh_microbenchmark}
\end{figure*}

We evaluate space overhead by comparing the file sizes of binaries protected by \ourproj{}-Base and \ourproj{}-EHProtect. As shown in Table~\ref{tab:loki_time}, enabling ABI-Compliant EH Shadowing increases the average binary size, which varies from 37.02 KB to 37.53 KB, with an average growth of 37.36 KB. 
Relative to the \ourproj{}-Base version, this represents an average size overhead of 66.26\% (ranging from 12.41\% to 113.14\% depending on the original binary size).
This increase consists of three components: 
(1) the obfuscated implementations of \texttt{EHInterceptor} and the relocated \texttt{LSHandler}, which are largely program-independent and therefore introduce a near-constant size overhead; 
(2) encrypted language-specific data (LSData), whose size scales with the number of protected functions and the complexity of program data structures; and 
(3) ABI-compliant shadow unwind codes.
Given the scale of modern software systems, we consider this space overhead negligible in exchange for substantially improved resistance to static analysis.

We measure the performance overhead by comparing the execution time between \ourproj{}-Base and \ourproj{}-EHProtect when exceptions are actually thrown. 
To this end, we develop a micro-benchmark suite designed to stress-test the virtual unwinding mechanism.
To ensure comprehensive coverage of C++ exception handling scenarios, we designed four representative test cases: 
(1) \emph{Primitive Type Propagation}, where we throw simple types (e.g., integers) to measure the basic cost of virtual stack walking without any object processing, thereby isolating the pure latency of the virtualization engine's control flow redirection; 
(2) \emph{Standard Exception Dispatch}, which throws \texttt{std::exception} instances to evaluate the cost of checking type information (RTTI) to match \texttt{catch} handlers, requiring the VM to dynamically parse encrypted metadata to verify type compatibility; 
(3) \emph{Stack Unwinding with RAII Cleanup}, which adds objects (e.g., \texttt{std::vector} standard containers) to stack frames to force the virtual unwinder to run destructors at each level, specifically measuring the overhead of the \emph{Local Unwind} phase where resource cleanup must be orchestrated before resuming traversal; 
and (4) \emph{Cascading Re-throw}, where an exception is caught and immediately re-thrown at every stack level to test how efficiently the virtual unwinding mechanism saves and restores exception state, stressing the context preservation logic during frequent mode transitions.
For each scenario, we employed the call stack nesting depth as the independent variable.
Specifically, we constructed a chain of non-inlined function calls ranging from depth 0 to 7; the exception is triggered in the deepest leaf function and propagated to the root caller. 
This setup forces the virtual unwinder to traverse an exact number of stack frames, allowing us to precisely quantify the per-frame processing cost.

The results are shown in Figure~\ref{fig:result_eh_microbenchmark}. The figure indicates that \ourproj{}-EHProtect incurs a performance overhead that increases approximately linearly with stack depth. 
This behavior aligns with our design expectations: deeper call stacks require additional unwinding operations within the virtual machine, and the execution of obfuscated code is inherently slower than native execution.

Despite this overhead, the absolute cost remains modest. 
Even in the worst case, handling an exception with a stack depth of seven takes approximately 150~\textmu s. 
Moreover, exception handling is not on the critical execution path of most software systems and occurs relatively infrequently in practice.
We therefore consider the performance impact of EH shadowing on real-world applications to be negligible.

\section{Related Work}
\label{sec:related}

\noindent\textbf{Software Obfuscation and Deobfuscation.}
Software obfuscation, formally categorized by Collberg et al.~\cite{collberg1997taxonomy, collberg1998manufacturing}, relies on transformations that increase the complexity of program logic to impede reverse engineering. Among these, VM-based obfuscation (or code virtualization) is widely regarded as one of the most resilient techniques~\cite{banescu2018tutorial, li2022chosen}.
Implementations typically target three distinct levels. 
Binary-level obfuscators operate directly on compiler-generated executables, a category that includes industry standards like VMProtect~\cite{vmprotect}, Code Virtualizer~\cite{codevirtualizer}, Themida~\cite{themida} and Obsidium~\cite{obsidium}, as well as academic prototypes like DynOpVm~\cite{cheng2019dynopvm}, which is developed based on ReWolf~\cite{rewolf}.
In academia, compiler-based approaches have flourished since the Obfuscator-LLVM~\cite{junod2015obfuscator}. Following this work, numerous LLVM-based obfuscation methods have been proposed to integrate directly into the compilation pipeline~\cite{jung2019fuzzification, xiao2023xvmp}. Advanced hardening techniques, such as Loki~\cite{schloegel2022loki}, further enhance this model by employing Mixed Boolean-Arithmetic (MBA) and synthesized superoperators. 
Finally, source-to-source obfuscators transform code prior to compilation. Tigress~\cite{tigress}, provides comprehensive obfuscation capabilities for the C language, including code virtualization.
\ourproj{} aligns with the binary-level obfuscators to maintain independence from build systems and provide flexible transformation pass interfaces for future work.

In parallel, automated deobfuscation has evolved into a sophisticated arms race. Early generic approaches relied on pattern matching or heuristics to locate VM structures~\cite{rolles2009unpacking, kinder2012towards, guillot2010automatic, sharif2009automatic, garba2019saturn}. 
As obfuscation complexity increased, researchers adopted dynamic binary instrumentation and symbolic execution to simplify execution traces~\cite{coogan2011deobfuscation, raber2013virtual, yadegari2015symbolic, yadegari2015generic, salwan2018symbolic, kalysch2017vmattack, xu2018vmhunt}. To address the path explosion challenges~\cite{banescu2016code}, synthesis-based methods such as Syntia~\cite{blazytko2017syntia}, Xyntia~\cite{menguy2021search}, and QSynth~\cite{david2020qsynth} were proposed to reconstruct instruction semantics by treating virtual handlers as black boxes.
More recently, inference-based attacks have emerged: Li et al.~\cite{li2022chosen} proposed the Chosen-Instruction Attack (CIA) to infer mapping rules via input-output analysis, while Luo et al.~\cite{luo2023reverse} introduced interpreter semantics testing to reverse engineer obfuscated Lua bytecode. Furthermore, Zhang et al.~\cite{zhang2025inspecting} demonstrated that incorporating knowledge of VM diversification can significantly enhance the effectiveness of existing tools like Virtual-Deobfuscator, VMHunt and Syntia.

However, existing obfuscation and deobfuscation research focus on control and data flow complexity, overlooking the structural information leaked by the exception handling metadata. \ourproj{} addresses this gap by offering the \textbf{ABI-Compliant EH Shadowing} mechanism that secures EH semantics.

\noindent\textbf{Exception Handling in Security Analysis.}
Exception-handling (EH) information, mandated by modern ABIs for stack unwinding, has become a reliable data source for binary analysis. The \texttt{.eh\_frame} section in ELF binaries and the \texttt{.pdata} in PE files contain precise mappings between code ranges and stack frame layouts. 
Pang et al.~\cite{pang2021towards} demonstrated that this EH metadata can be optimally used to detect function entry with high accuracy, surpassing traditional pattern-based heuristics. 
Furthermore, Duta et al.~\cite{duta2023let} revealed that EH metadata can be exploited for control flow hijacking attacks.
Modern reverse engineering tools like IDA Pro~\cite{IDAPro9EHparsing} and Ghidra~\cite{Ghidra} also extensively utilize EH metadata and language-specific data (LSData) to reconstruct function boundaries and control flow graphs (CFGs). 
Recognizing this vulnerability, recent works such as OCFI~\cite{pang2023ocfi, zhang2025posing} have proposed protecting EH metadata at the binary level to obstruct function entry detection. 
However, since these methods only make limited modifications to binary files, they can only partially protect the \emph{Global Section} and cannot protect the \emph{Local Section}.
Since the OS runtime requires plaintext access to EH metadata to perform stack unwinding correctly, existing VM-based obfuscators are forced to preserve it in its original form. This exposes static anchors that attackers can leverage to identify obfuscated functions and infer their internal structure. \ourproj{} addresses this limitation by introducing an exception-aware virtualization mechanism that secures EH metadata while maintaining ABI compatibility.

\section{Conclusion}
\label{sec:conclusion}

This paper presented \ourproj{}, a virtualization-based binary obfuscation framework that provides end-to-end protection for both executable code and exception-handling semantics. 
We identified exception-handling metadata as a critical yet underprotected attack surface in existing VM-based obfuscators and showed how it can be exploited for reverse engineering.
To address this gap, we proposed ABI-Compliant EH Shadowing, an exception-aware protection mechanism that preserves ABI compatibility while eliminating static exception metadata leakage by securely redirecting exception handling into a protected virtual machine. Our implementation and evaluation demonstrate that \ourproj{} preserves semantic correctness, disrupts automated reverse-engineering tools, and incurs negligible space overhead with modest runtime cost.
These results show that strong protection of exception handling is feasible in practice and can be integrated into VM-based obfuscation without sacrificing correctness or compatibility.

\bibliographystyle{ACM-Reference-Format}
\bibliography{cite_formatted}

\end{document}